\documentclass[aps,preprint]{revtex4}
\usepackage{amsmath}
\usepackage{amssymb}
\usepackage{graphicx}

\newcommand{\ds}{\displaystyle}

\begin{document}

\title{Correlation functions and ordering in a quasi-one dimensional system
of hard disks from the exact canonical partition function}
\author{V.M.~Pergamenshchik$^{1,2}$, T.M. Bryk$^{3}$, and A.D. Trokhymchuk$%
^{3}$}
\email{adt@icmp.lviv.ua}
\affiliation{$^{1}$Institute of Physics, National Academy of Sciences of Ukraine,
prospekt Nauky, 46, Kyiv 03039, Ukraine \\
$^{2}$Center for Theoretical Physics, Polish Academy of Sciences, Warsaw,
Poland \\
$^{3}$Institute for Condensed Matter Physics, National Academy of Sciences
of Ukraine, 1 Svientsitskii Str., Lviv 79011, Ukraine}
\date{\today }

\begin{abstract}
The analytical canonical $NLT$ partition function of a quasi-one dimensional (q1D)
system of hard disks [J. Chem Phys. \textbf{153}, 144111 (2020)] provides a
direct analytical description of the thermodynamics and ordering in this
system (a pore) as a function of the linear density $N/L$ both for finite
and infinite systems. It is alternative to the well-developed transfer
matrix method based on the isobaric $NPT$ ensemble. We derive the analytical
formulae for the actual distance dependence of the translational pair
distribution function (PDF) and the DF of distances between next neighbor
disks, and then demonstrate their use by calculating the translational order
in the pore. In all cases, the order is found to be of a short range and to
exponentially decay with the disks' separation. The correlation length
presented for the whole range of the q1D pore widths and different densities shows
a non-monotonic dependence with a maximum at the density $N/L=1$ and tends
to the 1D value for a vanishing pore width. The results
indicate a special role of this density when the pore length $L$ is equal
exactly to $N$ disk diameters. Considering orientational order, we show that derivation of 
the traditional PDF of transverse disks' coordinates reduces to the
eigenstate problem of the standard transfer matrix method and address the
difference between this last and our approach. We argue that the exponential 
decay of the transverse PDF reflects the absence of the microscopic
translational order and that a macroscopic orientational order can be
described separately. We introduce an orientational order parameter that
accounts for the local quality of the crystalline zigzag formed by disks and
is related to its structural element. This parameter is maximum in the
densely packed state and vanishes for a very low density. Due to specific
quasi-one dimensional geometry, this macroscopic order parameter is constant
along the pore so that the macroscopic orientational order is of a liquid
crystalline type.
\end{abstract}

\maketitle

\bigskip


\bigskip

\section{Introduction}

Macroscopic bodies consist of so many molecules that their number is often
referred to as infinite. As such, the statistical description of
many-particles bodies must deal with many, even infinite number of degrees
of freedom and as many integrals. As this limit can be studied only
theoretically, analytical results and particularly exact ones are of great
importance. To solve a statistical mechanical problem implies to reduce the
problem of calculation of its partition function (PF) and pair correlation
functions to a finite number of dimensions, finite number of integrals and
other mathematical actions. This is most often a task impossible and we try
to learn the physics of many-body system and develop the appropriate
mathematical tools by studying simplified models. In particular, a strong
simplification can be achieved by considering geometries with reduced
dimensionality and, in particular, one-dimensional (1D). A great number of
1D models considered in the last century and summarized in the book \cite%
{Lieb} has proved to be very usefully related to the physics in two and
three dimensions. In the theory of liquids modeling molecules as hard
spheres, the distinguished example of the 1D physics is the exact solution
for the PF of \ a 1D gas of hard core molecules, now known as Tonks' gas 
\cite{Tonks}.

The 1D Tonks gas is much simpler than any 2D system, nevertheless Tonks'
solution has become the analytical platform for further expansion into the
world of 2D HD systems via moving to certain q1D models. The simplest q1D
system is such that each disk can touch no more than one next neighbor from
both sides (the so-called single-file system); the width of such q1D pore
must be below ($\sqrt{3}/2+1$) times HD diameter. 
The analytical theory
of HDs in q1D pore was first considered by Wojciechovski et al \cite{Wojc}
for a system periodically replicated in the transverse direction. Later
Kofke and Post \cite{Kofke} proposed an approach that enables one to
consider HDs in a q1D pore in the thermodynamic limit of infinite number of
disks. This theory has become the main tool in studying a q1D one-file
system and eventually developed into the powerful transfer matrix method
(TMM) \cite{Varga,Gurin2013,Godfrey2014,Godfrey2015,Robinson,Hu,Comment}.
This method proved to be amenable to further development and generalization
to more complex q1D systems with a higher width and more next neighbors for
each disk \cite{Gurin2015,Godfrey2015,Robinson,Hu,Zarif}. Thanks to the
analytical methods, nowadays HDs in the q1D geometry have been intensively
used as a model glass former to study glass transitions and HDs' dynamics 
\cite{Robinson,Yamchi,Fu,Hicks,A i T, Godfrey2020,Zarif}. The new interest
has been brought about by the studies of actual physical ultra cold systems
such as Bose-Einstein condensates created in practically 1D or q1D
electromagnetic traps \cite{BEC}. Although mathematically quantum and
classical gases are very different, the classical 1D and q1D models can
provide some technical and even physical insight.

The transfer matrix method reduces the finding of the PF of a q1D
many-particle system to the eigenstate problem for certain linear operator,
the transfer matrix (TM), which amounts to solving an integral equation. In
general however the integral equation cannot be solved analytically \cite%
{Kamen,Varga}. The peculiarity of this method is that it is essentially
related to the pressure-based $NPT$ ensemble which does not directly predict
pressure as a function of system's width $D$ and length $L$: the solution of
the integral equation is first parametrized by the pressure $P$ and then one
finds a linear density $\rho =N/L$ which corresponds to this $P.$ Recently
one of us derived the exact analytical canonical $NLT$ PF of a q1D HD
one-file system (from now on q1D implies also one-file system) both for a
finite number of disks $N$ and in the thermodynamic limit \cite{JCP2020}. As
a result, finding the thermodynamic properties of a q1D HD system for given $%
L$ and $D$ is reduced to solving single transcendental equation which can be
easily done numerically. The PF, pressure along and across the pore,
distribution of the contact distances between neighboring HDs along the
pore, and distribution of HD centers across the pore are found analytically.
In this paper we derive and employ another fundamental thermodynamic
quantities, the pair distribution functions (PDF).

The corollary of the TMM is that if it is applicable to a given system, then
its correlation functions decay exponentially and the order is of a short
range. Solving the eigenstate problem in the TMM directly gives the leading
correlation length that describes the correlations between the disks'
transverse coordinates $y$'$s.$ This PDF $\,\left\langle y_{i}y_{i+n}\right\rangle\,$ is a function of the difference $\,n\,$ between the $\,y\,$ coordinates of two disks, numbered $\,i\,$ and $\,i+n\,$. For the longitudinal order, the TMM gives the subleading correlation length that describes the correlation between two pairs of neighbor disks separated by the order
number \cite{Godfrey2015}. At the same time, the most important correlation
functions are those that are functions of the actual distance $\,R\,$ between
disks along the system.
 The analytical formula for such PDF, $g_{1}(R),$ is known only for a 1D Tonks gas \cite{Frenkel,Yukhnovski,Santos}. Finding
the PDF $\,g(R)\,$ for a q1D system in the framework of the TMM is
possible by means of the following nontrivial numerical procedures: either
by inverse Fourier transform of the structure factor obtained from the
eigenstates of the TM \cite{Godfrey2015}, or by first planting the system's
configuration from the TM eigenstates and then computing the PDF from these
planted configurations \cite{Comment}. The main goal of this paper is to
develop an alternative, analytical approach to the PDFs of a q1D HD system based on the $NLT$ ensemble and demonstrate its implementation.

From the analytical canonical PF of q1D HD system~\cite{JCP2020}, we derive a formula for the translational PDF $\,g(R)\,$ which requires computing a few integrals
and can be straightforwardly implemented numerically. We also derive the PDF 
$\,g_{1}(R)\,$ for the distance $R$ between next neighbors. Both PDFs are
presented for an infinite system, but the canonical PF allows one to obtain
the formulas for finite systems, too. The PDFs are derived directly from the \textbf{canonical} 
PF without employing the standard Laplace' transform used for the derivation
of the PDF for a 1D Tonks gas \cite{Yukhnovski,Santos}. The method is first
demonstrated by \textbf{deriving/application} to the PDF of a 1D Tonks gas. The derived formulae
then used to obtain the translational PDF $g(R)$ of a q1D HD system and its correlation length
for a wide range of the pore widths and densities. In all cases, the order
is found to be of a short range and to exponentially decay with the disks'
separation. The correlation length presented for a wide range of the pore
widths and different densities shows a non-monotonic dependence with a
maximum exactly at the density $N/L=1$ and, for vanishing pore width, tends
to the 1D value of a Tonks gas.

Considering orientational order, we first show that, in the $NLT$ ensemble,
the PDF $\left\langle y_{i}y_{i+n}\right\rangle $ can be derived by the TMM
with the standard TM in which, however, the pressure parameterizing this TM,
is already known. We also address other differences between our and standard
TM approach. Then we discuss the PDF $\left\langle y_{i}y_{i+n}\right\rangle 
$ and show that it describes the orientational order at the microscopic
level. We argue that the exponential decrease of the transverse PDF reflects
the absence of the microscopic translational order and that the
orientational order can also be addressed at the macroscopic level by
analogy with the nematic order described by the macroscopic director. We
introduce the macroscopic order parameter for the zigzag disks' arrangement
in the pore which is related to the zigzag's structural element consisting
of two bonds between three neighboring disks. This order parameter is shown
to assume the maximum value $1$ for the dense packing and eventually
vanishes for low densities. The order parameter describes the local
macroscopic order, but is homogeneous and does not change along the
system indicating that the macroscopic liquid crystalline order in a q1D
system is of a long range.

The canonical PF and the methods of its calculations are introduced in
Sec.II, and then, in Sec. III, the formulas for PDFs $g(R)$ and $g_{1}(R)$
are derived. In Sec.IV\,, these formulae are used to study the
translational order and the correlation functions are presented. In Sec.V,
we consider the orientational order, the relation between our and the
standard TMM approach, and introduce the macroscopic order parameter. 
The final Sec.VI is devoted to the discussion of the role of density $N/L=1\,$, the relation between defects of the zigzag arrangement and the
decay of the microscopic transverse order, and of the macroscopic
orientational order parameter.

\section{Exact canonical partition function of a finite and infinite $\rm{q1D}$ HD system}

Consider a pore of the full width $D$ and finite length $L$ filled with a
finite number $N$ of HDs of diameter $d=1$. All lengths will be measured in
HD diameters. The reduced width $\Delta =(D-d)/d$, that gives the actual
pore width attainable to HD centers, in the single-file quasi 1D case ranges
from $0$ in the 1D case to the maximum $\sqrt{3}/2\approx 0.866.$ The $i$-th
disk has two coordinates, $x_{i}$ along and $y_{i}$ across the pore; $y$
varies in the range $-\Delta /2\leq y\leq \Delta /2$; the pore volume is $LD$%
. The vertical center-to-center distance between two neighbors, $\delta
y_{i}=y_{i+1}-y_{i},$ determines the contact distance $\sigma\,$ between
them along the pore: 
\begin{eqnarray}
\sigma (\delta y_{i}) &=&\min \left\vert
x_{i+1}(y_{i+1})-x_{i}(y_{i})\right\vert ,  \notag \\
\sigma (\delta y_{i}) &=&\sqrt{d^{2}-\delta y_{i}^{2}},  \label{sigma} \\
\sigma _{m} &=&\sqrt{d^{2}-\Delta ^{2}}\leq \sigma \leq d.  \notag
\end{eqnarray}

The minimum possible contact distance, $\sigma _{m}$, obtains for $\,\delta y=\pm \Delta\,$ when the two disks are in contact with the opposite walls.
Thus, each set of coordinates $\{y\}=y_{1}, y_{2},..., y_{N}$
determines the correspondent densely packed state of the total length $L^{\prime }\{y\}=\sum_{i=1}^{N-1}\sigma (\delta y_{i}),$ which we call condensate \cite{JCP2020}. The minimum condensate length is $\sigma_{m}(N-1)\,$, the maximum length can be as large as $(N-1)d$, but it cannot exceed $L-d\,$, i.e., $(N-1)\sigma_{m}<L^{\prime}\leq L_{\max}^{\prime}\,$ where $\,L_{\max}^{\prime}=\min [(N-1)d, L-d]\,$.

The exact PF of this q1D HD system is given by the following integral \cite{JCP2020}: 
\begin{equation}
Z=\Delta \int\nolimits_{-\infty }^{\infty }\frac{d\alpha }{N!}%
\int_{(N-1)\sigma _{m}}^{L_{m}^{\prime }}dL^{\prime }e^{i\alpha L^{\prime
}}(L-1-L^{\prime })^{N}\left( \int_{\sigma _{m}}^{1}\frac{dx}{\sqrt{1-x^{2}}}%
xe^{-i\alpha x}\right) ^{N-1}.  \label{Z}
\end{equation}%
It is convenient to rewrite this PF in the exponential form: 
\begin{equation}
Z=\frac{\Delta }{N!}\int_{(N-1)\sigma _{m}}^{L_{m}^{\prime }}dL^{\prime
}\int d\alpha e^{S},  \label{Z3}
\end{equation}%
where 
\begin{equation}
S=i\alpha L^{\prime }+N\ln (L-1-L^{\prime })+(N-1)\ln \left( \int_{\sigma
_{m}}^{1}\frac{dx}{\sqrt{1-x^{2}}}xe^{-i\alpha x}\right) .  \label{S}
\end{equation}

Equations~(\ref{Z3}) and (\ref{S}) give the PF in the general case of a q1D
HD system of any $N$ and $L.$ Of course, in the thermodynamic limit $N-1=N$
and $L-1=L,$ but we keep terms $O(1/N)$ as we also consider a finite $\,N$.

The integrand of $\,Z\,$ is a regular function of $\,\alpha \,$ so that the $%
\,\alpha -$integration contour, in particular, its central part that gives
the principal contribution to the integral, can be shifted while the ends
remain along the real axis. In the thermodynamic limit $N\rightarrow \infty
\,$, $\,N/L=\rho =const\,$, we can compute the PF (\ref{Z3}) by the steepest
descent method. In the limit $N\rightarrow \infty $ the integral (\ref{Z3})
is exactly determined by the saddle point which, for given $N\,$, $L\,$ and $\,\sigma_{m}\,$, is the stationary point of the function $\,S(i\alpha ,L^{\prime})\,$, Eq.~(\ref{S}). It is convenient to introduce real $\,a=i\alpha \,$ since $\,\alpha\,$ at the saddle point lies on the imaginary axis and the
integration contour has to be properly deformed. The two equations $\,\partial S/\partial a=\partial S/\partial L^{\prime }=0\,$ that determine the saddle point, can be reduced to the single equation for $\,a=a_{N}\,$ which reads: 
\begin{equation}
\frac{L}{N}-\frac{1}{a_{N}}=\frac{I^{\prime }(a_{N})}{I(a_{N})}\,,  \label{E}
\end{equation}
where 
\begin{eqnarray}
I(a_{N}) &=&\int_{\sigma _{m}}^{1}\frac{dx}{\sqrt{1-x^{2}}}x\exp (-a_{N}x),
\label{I} \\
&&  \notag \\
I^{\prime }(a_{N}) &=&\int_{\sigma _{m}}^{1}\frac{dx}{\sqrt{1-x^{2}}}%
x^{2}\exp (-a_{N}x)\,.  
\label{I1}
\end{eqnarray}
The solution $\,a_{N}\,$ of Eq.~(\ref{E}), which gives the total
longitudinal force $\,Ta_{N}\,$ and longitudinal pressure $\,P_{L}=Ta_{N}/D\,$, depends on the per disk pore length $\,L/N\,$
and, via $\sigma_{m}\,$, on the pore width $\,D\,$, and
fully determines the free energy. The free energy per disk,$\,F/N\,$, which therefore is the function of the pore length $\,L\,$, pore width $\,D\,$ and the temperature $\,T\,$, is $\,F(L,D,T)/N=-TS(a_{N})/N=-Ts_{N}\,$ where $\,s_{N}\,$ is system's per disk entropy\,: 
\begin{equation}
s_{N}=a_{N}\sigma _{N}+\ln \left( L-N\sigma _{N}\right) +\frac{N-1}{N}\ln
I(a_{N})\,,  \label{S1}
\end{equation}%
where $\sigma _{N}$ is the average value of the contact distance $\,\sigma\,$ in the condensate [i.e., average of $\,L^{\prime }/(N-1)\,$] \cite{JCP2020}%
: 
\begin{equation}
\sigma _{N}=\frac{L}{N}-\frac{1}{a_{N}}\,.  \label{sigN}
\end{equation}%
Finally, for $\,N\rightarrow \infty \,$, the PF can be cast in the two
equivalent forms: 
\begin{eqnarray}
Z_{\infty } &=&\frac{\varsigma _{N}\Delta }{N!}\exp (Ns_{N})  \label{ZZZ} \\
&=&\frac{\varsigma _{N}\Delta }{N!}(L-N\sigma
_{N})^{N}I(a_{N})^{N-1}e^{Na_{N}\sigma _{N}}.  \notag
\end{eqnarray}%
where $\varsigma _{N}$ is the prefacfor $\sim 1/\sqrt{N}$ originated from
the Gaussian integration along the steepest descent contour whose exact form
is of no need. In the 1D case, all $\sigma $'s are equal to $\,d\,$ and this
expression goes over into the Tonks PF $\,Z_{1D}\,$ up to the factor $\Delta
^{N}$ which in this case represents the independent transverse degrees of
freedom: $\,Z_{\infty }\rightarrow \Delta ^{N}Z_{1D}\,$ where 
\begin{equation}
Z_{1D}=\frac{1}{N!}\left( L-Nd\right) ^{N}\theta \left( L-Nd\right) \,.
\label{Z1D}
\end{equation}

Now consider the general case of a finite system. In what follows, the
number of HDs and the total length of a finite system are denoted as $\,n\,$ and $\,R\,$, respectively (instead of $\,N\,$ and $\,L\,$). The integral (\ref{Z})
can be transformed to the one along the real axis $\alpha \,$like that: 
\begin{eqnarray}
Z_{n,R} &=&\frac{\Delta }{n!}\int_{\ds{(n-1)\sigma _{m}}}^{%
\ds{L_{m}}}dL^{\prime }\left( R-1-L^{\prime }\right) ^{%
\ds{n}}  \notag \\ 
\label{Z1}
&& \\ 
&&\times \int_{-\infty }^{\infty }\frac{\ds{d\alpha }}{%
\ds{2\pi }}\left\vert I(i\alpha )\right\vert ^{\ds{n-1}%
}\cos \left[ L^{\prime }\alpha +(n-1)\varphi _{\alpha }\right] \,,  \notag
\end{eqnarray}%
where $L_{m}=\min (n-1,R-1)$ and 
\begin{eqnarray}
\varphi _{\alpha } &=&\arg I(i\alpha ),  \notag  \label{bII} \\
&& \\
I(i\alpha ) &=&\int_{\sigma _{m}}^{1}\frac{dx}{\sqrt{1-x^{2}}}xe^{%
\displaystyle{-i\alpha x}}\,.  \notag
\end{eqnarray}

Although the Gaussian approximation at the saddle point cannot give an exact
result for a system with finite number of disks, choosing the $\alpha\,$
integration contour passing through the saddle point provides the best
convergence of the integrals (which has been confirmed by numerically). Hence
to compute the PF we shift the central part of the $\,\alpha\,$ integration
contour downward and integrate over the real variable $t$ along the line $\,\alpha =-ia_{n}+t\,$ that crosses the imaginary axis at $\,\alpha =-ia_{n}\,$. The best choice for the shift $\,a_{n}\,$ is the root of the following modified equation (\ref{E}): 
\begin{equation}
\frac{R}{n}-\frac{n-1}{na_{n}}=\frac{I^{\prime }(a_{n})}{I(a_{n})}\,,
\label{En}
\end{equation}
whose rhs is defined in Eqs.~(\ref{I}) and (\ref{I1}). Then the PF $Z_{n,R}$  can be transformed like that: 
\begin{eqnarray}
Z_{n,R} &=&\frac{\Delta }{n!}\int_{\displaystyle{(n-1)\sigma_{m}}}^{%
\displaystyle{L_{m}}}dL^{\prime }e^{\displaystyle{a_{n}L^{\prime }}}\left(
R-1-L^{\prime }\right) ^{n}  \notag  \label{ZnR} \\
\\
&&\times \int_{-\infty }^{\infty }\frac{dt}{2\pi }\left(
I_{s}{}^{2}+I_{c}^{2}\right) {}^{(n-1)/2}\cos [L^{\prime }t+(n-1)\varphi ]\,.
\notag
\end{eqnarray}
where 
\begin{eqnarray}
I_{s}(t) &=&-\int_{\sigma _{m}}^{1}\frac{dx}{\sqrt{1-x^{2}}}xe^{\displaystyle%
{-a_{n}x}}\sin(tx)\,,  \notag  \label{IcIs} \\
\\
I_{c}(t) &=&\int_{\sigma _{m}}^{1}\frac{dx}{\sqrt{1-x^{2}}}xe^{\displaystyle{%
-a_{n}x}}\cos(tx)\,,  \notag
\end{eqnarray}
\begin{equation}
\varphi (t)=\arg \left( I_{c}+iI_{s}\right) =\left\{ 
\begin{array}{c}
\arctan \frac{I_{s}}{I_{c}},I_{c}>0, \\ 
\pi +\arctan \frac{I_{s}}{I_{c}},I_{c}<0,I_{s}>0, \\ 
-\pi +\arctan \frac{I_{s}}{I_{c}},I_{c}<0,I_{s}<0.%
\end{array}%
\right. .  \label{arg2}
\end{equation}
The density $\rho _{n}=n/R$ and the reduced pore width $\Delta\,$, which enter the integrals above via $\,\sigma_{m}\,$, fully determine the partition function $\,Z_{n,R}\,$ through Eqs.~(\ref{En})-(\ref{arg2}).

\section{Derivation of the translational PDF from the
canonical partition function}

The PDF\ as a function of separation $R$ is the probability to find particle
a distance $R$ from another particle whose coordinate $x_{0}$ is fixed, say
at $x_{0}=0.$ In a 1D HD system, the PDF $g(R)$ is usually found from the isobaric $\,NPT\,$ ensemble by means of Laplace's transform \cite{Yukhnovski,Santos}. Here
we derive $\,g(R)\,$ for a q1D HD systems directly from the PF $%
\,Z_{N}\{x_{i},y_{i}\}\,$ of the canonical $NLT$ ensemble.

The q1D PF $Z_{N}\{x_{i},y_{i}\}$ is a functional of the particles'
longitudinal $x$ coordinates and transverse $y$ coordinates. In the
particular case of a q1D system, the general formula for the PDF $g(R)$
equivalent to its definition is obtained from the canonical PF for the $N$
particle system by fixing the $x$ coordinate of $n$-th disk at $x_{n}=x$ and
then summing over all possible $n$ (the range of $n$ will be clarified later on)\,: 
\begin{equation}
g(R)=\frac{1}{\rho }\sum_{n=1}\frac{Z_{N}%
\{x_{0}=0,y_{0},x_{1},y_{1},...,x_{n}=R,y_{n},...,x_{N},y_{N}\}}{%
Z_{N}\{x_{0}=0,y_{0},x_{1},y_{1},...,x_{n},y_{n},...x_{N},y_{N}\}}\,.
\label{P(R)}
\end{equation}
Note that $y_{0}$ and $y_{n}$ are not fixed so that the particles $0$ and $n$
can move in the transverse direction. The PF in the nominator splits into a
product of two PFs, $Z_{n}$ for $n$ disks (of which $n-1$ free to move) in
the space $0<x_{k}<R,$ and $Z_{N-n,L-R}$ for $N-n$ moving disks in the space  $\,R<x_{k}<L-R-d/2\,$, Fig.1\,:
\begin{equation}
g(R)=\frac{1}{\rho }\sum_{n=1}\frac{Z_{n,R}Z_{N-n,L-R}}{Z_{N,L}}\,,
\label{P1}
\end{equation}
where $\rho =N/L$ is the linear density. Figure~1 demonstrates that the numbers of free disks, contact distances $\sigma\,$, and disk-wall distances $d/2$
have to be adjusted in each PF individually. As a result, the form of Eq.~(\ref{E}) that determines $a_{n}$, is also slightly modified.

\begin{figure}[tbp]
\includegraphics[width=0.75\textwidth]{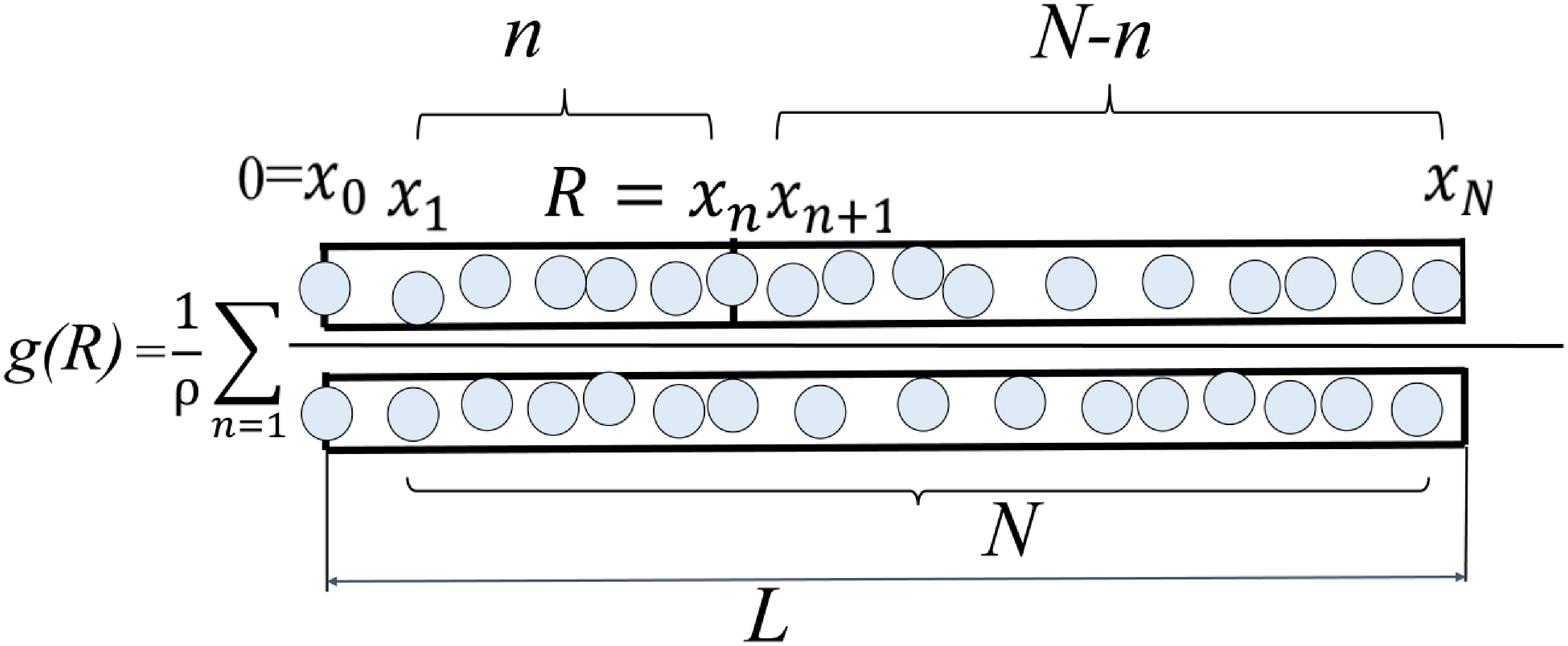}
\caption{Definition of the pair distribution function $\,g(R)\,\ $and the
three PFs. $Z_{n,R}$ is for $n-1$ free moving disks in the space $R,$ $%
Z_{L-R,N}$ is for $N-n$ free disks in the space $\,L-R\,$, and $Z_{N}$ is
for $N$ free disks in the space $\,L\,$.}
\label{gR}
\end{figure}

Consider first $Z_{n,R}$. We assume that $R>1;$ the case $R<1$, possible
only for $n=1,$ will be considered separately. In the system of size $R>1,$
there are $n-1$ freely moving HDs, $\int dy_{0}=\Delta ,$ $n$ contact
distances $\sigma $, and no walls keeping disks at the minimum distance $d/2\,$. Hence $(R-d-L^{\prime})^{n}$ in PF (\ref{ZnR}) has to be replaced by $\,(R-L^{\prime})^{n-1}$ and $I(i\alpha )^{n-1}$ by $I(i\alpha )^{n}\,$. Then the PF $Z_{n,R}(R)$ takes the form 
\begin{eqnarray}
Z_{n,R} &=&\frac{\Delta }{(n-1)!}\int_{n\sigma _{m}}^{L_{n,m}}dL^{\prime }e^{%
\displaystyle{a_{n}L^{\prime }}}\left( R-L^{\prime }\right) ^{n-1}  \notag
\label{ZZn} \\
\\
&&\times \int_{-\infty }^{\infty }\frac{dt}{2\pi }%
(I_{s}{}^{2}+I_{c}{}^{2})^{n/2}\cos [L^{\prime }t+n\varphi (t)]\,,  \notag
\end{eqnarray}
where $L_{n,m}=\min(n,R)\,$. The angle $\,\varphi(t)\,$ is defined in Eq.~(\ref{arg2}), $\,a_{n}\,$ is the root of equation (\ref{En}), and the relation between $\sigma_{n}$ and $a_{n}$ is just the properly modified Eq.~(\ref{sigN})\,, 
\begin{equation}
\sigma_{n}=\frac{R}{n}-\frac{n-1}{na_{n}}\,.  
\label{b}
\end{equation}

Next consider $Z_{N-n,L-R}$ in Eq.~(\ref{P1}), the PF for $N-n$ HDs in the range $R<x<L\,$. Here all disks are free to move, there are $N-n$ contact distances $\sigma\,$, and the singe wall at the pore end. Then the PF $\,Z_{N-n,L-R}(R)\,$ can be presented in the form 
\begin{eqnarray}
Z_{N-n,L-R} &=&\frac{1}{(N-n)!}\int_{\sigma _{m}}^{L_{N-n,m}}dl e^{%
\displaystyle{L^{\prime}a_{N-n}}}(L-R-1/2-L^{\prime })^{N-n}  \notag
\label{ZN-n} \\
\\
&&\times \int_{-\infty }^{\infty }\frac{dt}{2\pi }%
(I_{s}{}^{2}+I_{c}{}^{2})^{(N-n)/2}\cos [L^{\prime }t+(N-n)\varphi (t)]\,, 
\notag
\end{eqnarray}
where $\,L_{N-n, m}=\min(N-n,L-R-1/2)$ and $a_{N-n}$ is the root of the following modified Eq.~(\ref{E}):
\begin{equation}
\frac{L-R-1/2}{N-n}-\frac{1}{a_{N-n}}=\frac{I^{\prime }(a_{N-n})}{I(a_{N-n})}%
\,.  
\label{EN-n}
\end{equation}

At last, consider $Z_{N,L}$ in Eq.~(\ref{P1}) that is the PF for $N$ HDs in
the range $0<x<L.$ Here all disks are free to move, $\int dy_{0}=\Delta ,$
there are $N$ contact distances $\sigma $ and the singe wall at the pore
end. Then the PF $Z_{N,L}(R)$ can be presented in the form 
\begin{eqnarray}
Z_{N,L} &=&\frac{\Delta }{N!}\int_{N\sigma _{m}}^{L_{N,m}}dL^{\prime }e^{%
\displaystyle{a_{N}L^{\prime }}}(L-1/2-L^{\prime })^{N}  \notag  \label{ZNL}
\\
\\
&&\times \int_{-\infty }^{\infty }\frac{dt}{2\pi }%
(I_{s}{}^{2}+I_{c}{}^{2})^{N/2}\cos [L^{\prime }t+N\varphi (t)]\,,  \notag
\end{eqnarray}
where $L_{N,m}=\min (N,L-1/2)$ and $a_{N}$ is the root of the equation (\ref%
{E}). Making use of the PFs (\ref{ZZn}), (\ref{ZN-n}), and (\ref{ZNL}) in
the general formula (\ref{P1}) gives the PDF of a q1D HD system for finite $%
N\,$ and $L\,$.

The general result for $g(R)$ can be further simplified in the thermodynamic
limit $N\rightarrow \infty ,L\rightarrow \infty ,N/L=\rho =const\,$. This
case, usually considered the most important one, is presented in detail in
the next section. To illustrate our method of deriving the PDF directly from
the canonical $NLT$ PF we first derive the PDF for 1D Tonks' gas.

\subsection{PDF of a q1D HD system in the thermodynamic limit}

\subsubsection{PDF of an infinitely long 1D HD system (Tonks' gas)}

The PDF $g(R)$ for a 1D HD is given by the general formula (\ref{P1}) in which the three PFs are obtained from the Tonks' PF, Eq.~(\ref{Z1D})\,: 
\begin{equation}
g_{1D}(R)=\frac{1}{\rho }\sum_{n=1}\frac{N!|R-n|^{n-1}[L-R-n]^{N-n}}{%
(n-1)!(N-n)!(L-n)^{N}}\,\theta (R-n)\,.  
\label{g1D}
\end{equation}
In the limit $N\rightarrow \infty\,$, neglecting $\,O(n/N)\,$, one also has $\,(N-n)!\cong N!/N^{n}\,$ and 
\begin{eqnarray}
(L-R-n)^{N-n} &=&(L-N)^{N-n}\left( 1-\frac{R-n}{N(l_{N}-1)}\right) ^{N-n} 
\notag \\
&=&(L-N)^{N-n}\left[ \exp \left( -\frac{R-n}{l_{N}-1}\right) +O\left( \frac{n%
}{N}\right) \right]  \label{e1} \\
&\rightarrow &(L-N)^{N-n}\exp \left( -\frac{R-n}{l_{N}-1}\right)\,,  \notag
\end{eqnarray}
where $l_{N}=L/N=1/\rho\,$. Making use of these results in Eq.~(\ref{g1D}), one finally obtains 
\begin{equation}
g_{1D}(R)=\frac{1}{\rho }\sum_{n=1}\frac{|R-n|^{n-1} \exp\left(\displaystyle{%
\ -\frac{R-n}{l_{N}-1}}\right)}{(n-1)!(l_{N}-1)^{n}}\,\theta (R-n)\,,
\label{gg1D}
\end{equation}
which is the well-known PDF of 1D Tonks' gas \cite{Yukhnovski,Santos}.

\subsubsection{PDF of an infinitely long q1D HD system.}

In an infinitely long q1D HD system, the above thermodynamic limit result,
Eqs.~(\ref{Z3}) and (\ref{S}), is applicable both for $Z_{N,L}$ and $%
Z_{N-n,L-R}\,$ as the number of particles $N-n$ and volume $L-R$ are
infinite, but the PF $Z_{n,R}$ for the finite $n$ disk system has to be
found directly from the general formula (\ref{ZnR}) [or from the original form (\ref{Z1}) without the contour shift]. Adjusting Eqs.~(\ref{S1})-(\ref{ZZZ}) to the above PFs of interest, one has: 
\begin{eqnarray}
Z_{N-n,L-R} &=&\frac{\varsigma _{N-n}}{(N-n)!}[L-R-(N-n)\sigma
_{N-n}]^{N-n}\exp [(N-n)\widetilde{s}_{N-n}]\,,  \notag  \label{ZZ} \\
&& \\
Z_{N,L} &=&\frac{\varsigma _{N}\Delta }{N!}(L-N\sigma _{N})^{N}\exp (N%
\widetilde{s}_{N-n})\,.  \notag
\end{eqnarray}%
Here $\widetilde{s}_{N-n}=a_{N-n}\sigma _{N-n}+\ln I(a_{N-n})$ and $%
\widetilde{s}_{N}=a_{N}\sigma _{N}+\ln I(a_{N})$, where the pair $\sigma
_{N},a_{N}$ is determined by $l_{N}=L/N=1/\rho \,$ from the Es.~(\ref{E})
and (\ref{sigN}) and the pair $\sigma _{N-n},a_{N-n}$ by $l_{N-n}=(L-R)/(N-n)
$ from similar equations 
\begin{eqnarray}
l_{N-n}-\frac{1}{a_{N-n}} &=&\frac{I^{\prime }(a_{N-n})}{I(a_{N-n})}\,, 
\notag  \label{sigN-n} \\
&& \\
\sigma _{N-n} &=&l_{N-n}-\frac{1}{a_{N-n}}\,.  \notag
\end{eqnarray}
Substituting these expressions in the general formula (\ref{P1}) for $g(R)$
and taking into account that in the thermodynamic limit the preexponential
factors $\varsigma _{N}$ and $\varsigma _{N-n}$ are equal, we get: 
\begin{eqnarray}
g(R) &=&\frac{1}{\rho }\sum_{n=1}\frac{Z_{n,R}N![L-R-(N-n)\sigma
_{N-n}]^{N-n}}{(N-n)!(L-N\sigma _{N})^{N}}  \notag  \label{P2} \\
&& \\
&&\times \exp [N(\widetilde{s}_{N-n}-\widetilde{s}_{N})-n\widetilde{s}%
_{N-n}]\,.  \notag
\end{eqnarray}
Now we find $\widetilde{s}_{N-n}$ by expanding about $\widetilde{s}_{N}$ and
using the smallness of $\ n/N.$ First, up to $O(n/R),$ one has $N(\widetilde{%
s}_{N-n}-\widetilde{s}_{N})\cong N\left( \partial \widetilde{s}_{N}/\partial
l_{N}\right) (l_{N-n}-l_{N})\,$, where 
\begin{eqnarray}
l_{N-n}-l_{N} &=&\frac{L-R}{N-n}-\frac{L}{N}  \notag  \label{del l} \\
&& \\
&=&\frac{R-l_{N}n}{N}\left[ 1+O(n/L)\right] \,.  
\notag
\end{eqnarray}
The $\,l_{N}\,$ derivative obtains regarding (\ref{E}) and (\ref{sigN})\,: 
\begin{equation}
\frac{\partial\widetilde{s}_{N}}{\partial l_{N}}=
\frac{1}{a_{N}}\frac{\partial a_{N}}{\partial l_{N}} + a_{N} = a_{N}\frac{\partial\sigma_{N}}{\partial l_{N}}\,.
\label{dsN}
\end{equation}
Then one expands $\widetilde{s}_{N-n}$ about $\widetilde{s}_{N}$ regarding (\ref{del l})\,: 
\begin{eqnarray}
n\widetilde{s}_{N-n} &\cong &n\widetilde{s}_{N} + 
n\frac{\partial\widetilde{s}_{N}}{\partial l_{N}}(l_{N-n}-l_{N})  \label{Ns1} 
\\
&=&n\widetilde{s}_{N}+O(n/N)\,,  
\notag
\end{eqnarray}
to finally obtain 
\begin{equation}
N(\widetilde{s}_{N-n}-\widetilde{s}_{N})\cong - 
a_{N}\frac{\partial\sigma_{N}}{\partial l_{N}}( R-nl_{N})\,.  
\label{Nds}
\end{equation}
Next we show that the $N-n$ th power of the ratio in (\ref{P2}) gives rise to an exponential: 
\begin{eqnarray}
&&\left[ \frac{L-R-(N-n)\sigma _{N-n}}{L-N\sigma _{N}}\right] ^{N-n}  \notag
\\
&=&\left( \frac{l_{N}-\sigma _{N-n}}{l_{N}-\sigma _{N}}\right) ^{N-n}\left(
1-\frac{R-n\sigma _{N-n}}{L-N\sigma _{N-n}}\right) ^{N-n}  \label{exp} \\
&\cong &\left( \frac{l_{N}-\sigma _{N-n}}{l_{N}-\sigma _{N}}\right)
^{N-n}\exp \left( -\frac{R-n\sigma _{N}}{l_{N}-\sigma _{N}}\right) .  \notag
\end{eqnarray}
In turn, the first factor in the last line can also be reduced to an
exponential whose exponent cancels out the one in Eq.~(\ref{Nds})\,:
\begin{eqnarray}
\left( \frac{l_{N}-\sigma _{N-n}}{l_{N}-\sigma _{N}}\right) ^{N-n} &=&\left(
1+\frac{\sigma _{N}-\sigma _{N-n}}{l_{N}-\sigma _{N}}\right) ^{N-n}\cong  
\notag  \label{exp2} \\
&& \\
\left[1+\frac{a_{N}}{N}\frac{\partial\sigma_{N}}{\partial l_{N}}\right]^{N-n} 
&\cong & 
\exp \left[ a_{N}(R-nl_{N})\frac{\partial\sigma_{N}}{\partial l_{N}} \right] \,.  
\notag
\end{eqnarray}
Making use of the results (\ref{Nds})-(\ref{exp2}) in formula (\ref{P2}), after some straightforward algebra and convenient rescaling, we obtain the PDF in the final form: 
\begin{equation}
g(R)=\frac{1}{\rho }\sum_{n=1}^{n_{\max }}\frac{|R-n\sigma _{n}|^{n-1}\exp
\left\{ -\frac{\displaystyle{R-n\sigma _{N}}}{\displaystyle{l_{N}-\sigma _{N}%
}}+n\left[ a_{n}\sigma _{n}-a_{N}\sigma _{N}+\ln \frac{\displaystyle{I(a_{n})%
}}{\displaystyle{I(a_{N})}}\right] \right\} }{(n-1)!(l_{N}-\sigma _{N})^{n}}%
\,J_{n}(R)\,.  
\label{PDF}
\end{equation}
Here $\,J_{n}(R)\,$ is the following integral: 
\begin{eqnarray}
J_{n}(R) &=&n\int_{\ds{\sigma _{m}}}^{\ds{l_{m}}}dle^{na_{n}(l-\sigma _{n})}\left( 
\frac{R/n-l}{|R/n-\sigma _{n}|}\right)^{\ds{n-1}} \label{Jn} 
\notag \\
&& \\
&&\times \int_{-\infty }^{\infty }\frac{dt}{2\pi }\left[ \frac{%
I_{c}(t)^{2}+I_{s}(t)^{2}}{I(a_{n})^{2}}\right] ^{\ds{n/2}}\cos [n(lt+\varphi )]\,, 
\notag  
\end{eqnarray}
where $\,I_{c}(t), I_{s}(t), \varphi(t)\,$ and $\,a_{n}, \sigma_{n}\,$ are given in
Eqs.~(\ref{IcIs}), (\ref{arg2}) and Eqs.~(\ref{En}), (\ref{b}),
respectively. Deriving Eqs.~(\ref{PDF}) and (\ref{Jn}), we changed from the
variable $L^{\prime }$\ to $l=L^{\prime }/n$\ so that the upper $l$\
integration limit is now $l_{m}=\min (1,R/n).$\ To avoid dealing with
extremely small quantities and extremely fast oscillations, we made the
following convenient rescaling: we divided $R/n-l$ by $|R/n-\sigma _{n}|$\
and, to compensate, introduced the factor $|R-n\sigma _{n}|^{n-1};$\
similarly, the factor $\,\exp [na_{n}\sigma _{n}+n\ln I(a_{n})]\,$ compensates
for the denominator $I(a_{n})^{{n}}\,$ and $\,\exp[-na_{n}\sigma_{n}]\,$ in the
integrand.

The maximum 
$n_{\max }$ in summation of Eq.~({\ref{PDF}}) is the maximum number of disks
at close contact which can be put in the space between the particle fixed at $\,x=0\,$ and the point $\,x=R\,$: 
\begin{equation}
n_{\max }(R)=\frac{R-\mathrm{mod}(R,\sigma_{m})}{\sigma_{m}}\,.
\label{n max}
\end{equation}
Note that the expression for $\,g(R)\,$ appears to be considerably simpler if no contour shift and rescaling have been applied:
\begin{eqnarray}
g(R) &=&\frac{1}{\rho }\sum_{{n=1}}^{{n_{\max}}}\frac{n\exp \left(\ds{-\frac{R-n\sigma_{N}}{l_{N}-\sigma_{N}}}\right) }{(n-1)!(l_{N}-\sigma _{N})^{n}} 
\notag \\
&&  \label{gsimpl} \\
&&\times\int_{\sigma_{m}}^{l_{m}}dl\left( R/n-l\right) ^{n-1}\int_{-\infty
}^{\infty }\frac{d\alpha }{2\pi }\left\vert I(i\alpha )\right\vert
^{n/2}\cos [n(l\alpha +\varphi_{\alpha })]\,,  
\notag
\end{eqnarray}
where $\,I(i\alpha)\,$ and $\,\varphi_{\alpha}\,$ are defined in (\ref{bII}). But
the formulae (\ref{PDF}) and (\ref{Jn}) actually provide a much better
convergence and much simpler numericals.

\subsubsection{The 1D limit}

It is important to see how the results obtained for a q1D HD system behave
approaching a 1D HD system, i.e., in the limit $D\rightarrow 0$ when $\Delta
\rightarrow 0\,$, and $\sigma _{m},\sigma _{n},\sigma _{N}\rightarrow 1\,$.
To this end, we first estimate the $x$ integrals in this limit: 
\begin{eqnarray}
I(a) &=&e^{-a}\Delta +O(\Delta ^{2})\,,  \notag \\
I_{c} &=&e^{-a}\Delta \cos t+O(\Delta ^{2})\,,  \label{I0} \\
I_{s} &=&e^{-a}\Delta \sin t+O(\Delta ^{2})\,.  \notag
\end{eqnarray}%
As a result, 
\begin{eqnarray}
\varphi  &\rightarrow &-t\,,  \notag \\
&&  \notag \\
\frac{I_{c}^{2}+I_{s}^{2}}{I(a_{n})^{2}} &\rightarrow &1\,,  \notag \\
&&  \notag \\
\int_{-\infty }^{\infty }\frac{dt}{2\pi }\left[ \frac{I_{c}^{2}+I_{s}^{2}}{%
I(a_{n})^{2}}\right] ^{n/2}\cos [n(lt+\varphi )] &\rightarrow &\delta
(l-1)\,,  \label{delta(l-1)} \\
&&  \notag \\
J_{n} &\rightarrow &1\,.  \notag
\end{eqnarray}%
We see that in the 1D limit, the $g(R)$ (\ref{PDF}) goes over into the Tonks 
$g_{1D}(R)$, Eq.~(\ref{gg1D}).

\subsubsection{Probability to find next neighbor at a distance $R$ $(n=1)$.}

The term with $n=1$ in the PDF $g(R)$ is proportional to the probability $g_{1}(R)$ to have disk's next neighbour at a distance $R\,$. Here we derive
this important quantity for all $R$ larger than the minimum contact distance 
$\sigma _{m}$ for an infinitely long q1D HD system. In this case the shift
of the $\,\alpha \,$ contour to $\,a_{1}\,$ does not work because the
Eq.~(\ref{En}) breaks down, and we choose the shift $\,a_{N}\,$.
Regarding the equalities $\sigma _{N-1}=\sigma _{N}$ and $\widetilde{s}%
_{N-1}=\widetilde{s}_{N}$ and retaining only the $\,R\,$ dependent terms in Eq.~(\ref{PDF}), one has: 
\begin{equation}
g_{1}(R)\propto Z_{1,R}\exp \left( -\frac{R-\sigma _{N}}{l_{N}-\sigma _{N}}%
\right) ,  \label{g1}
\end{equation}
where PF $\,Z_{1,R}\,$, computed directly from its definition, has the form 
\begin{equation}
Z_{1,R}=\left\{ 
\begin{array}{c}
\displaystyle{2\int_{\sqrt{1-R^{2}}}^{\Delta }dy_{0}\int_{0}^{y_{0}-\sqrt{%
1-R^{2}}}dy}=\left( \Delta -\sqrt{1-R^{2}}\right) ^{2},\,\,R<1\,, \\ 
\\ 
\Delta ^{2},\,\,R\geq 1\,.%
\end{array}%
\right. .  
\label{Z1R}
\end{equation}
Normalizing on unity, one finally obtains: 
\begin{equation}
g_{1}(R)=\frac{Z_{1,R}\exp \left( \displaystyle{-\frac{R-\sigma _{N}}{%
l_{N}-\sigma _{N}}}\right) }{\displaystyle{\int_{\sigma _{m}}^{\infty
}dRZ_{1,R}\exp \left( -\frac{R-\sigma _{N}}{l_{N}-\sigma _{N}}\right) }}\,.
\label{gg1}
\end{equation}

\section{Translational order}

Figure 2 presents the DF $\,g_1(R)\,$ between next neighbor disks
obtained from Eq.~(\ref{gg1}) for a set of linear densities $\,\rho_{N}=N/L\,$ and two reduced pore widths $\,\Delta\,$. The density $\,\rho_{N}\,$ determines $\,a_{N}\,$ via simple transcendental Eq.~(\ref{E}) in which $\,\Delta\,$
enters via maximum contact distance $\,\sigma_{m}\,$, Eq.~(\ref{sigma}), and $\,\sigma_{N}\,$ is
given in Eq.~(\ref{sigN}). The sharp peak at $\,R=1\,$ is present at all densities including very high, but in this case its hight is incomparable with the second peak centered at the average interdisk spacing $\,l_{N}=1/\rho_{N}\,$. The second peak appears and strengthens as density becomes higher and
higher. The concentration of spacings $\,R\,$ at the average distance
indicates a high order in the $\,R$ direction. For high densities close to
the dense packing, this also implies a high overall zigzag order since $%
\,R\cong l_{N}\,$ approaches the minimum separation $\,\sigma _{m}\,$ for
which disks stay very close to the walls. In contrast, the fact that there
is a high peak at $\,R\approx 1\,$ which is particularly pronounced for the
density $\,\rho =1\,$ with $\,l_{N}\,$ exactly one shows that the ordering
at this density is not necessarily related with a zigzag type order. We
shall give this aspect a more consideration later on as the peculiarity of
separation $\,R=1\,$ and density $\,\rho =1\,$ will get additional
indications. Right now we would like only to explain the very reason for the
cusp at $\,R=1\,$ whose presence at $\,g_{1}(R)\,$ and PDF $\,g(x)\,$ has
been well known \cite{Godfrey2015,JCP2020,WE,Comment}. The explicit
analytical form of $g_{1}(R)\,$, Eq.~(\ref{Z1R}), shows that the increase of
the transverse free path of a disk at $\,R\,$ with $\,R\,$ abruptly stops at
its maximum constant value $\,\Delta \,$ as $\,R\,$ attains and exceeds the
value of the disk diameter $d$ $(=1)\,$. 
\begin{figure}[tbp]
\includegraphics[width=0.475\textwidth]{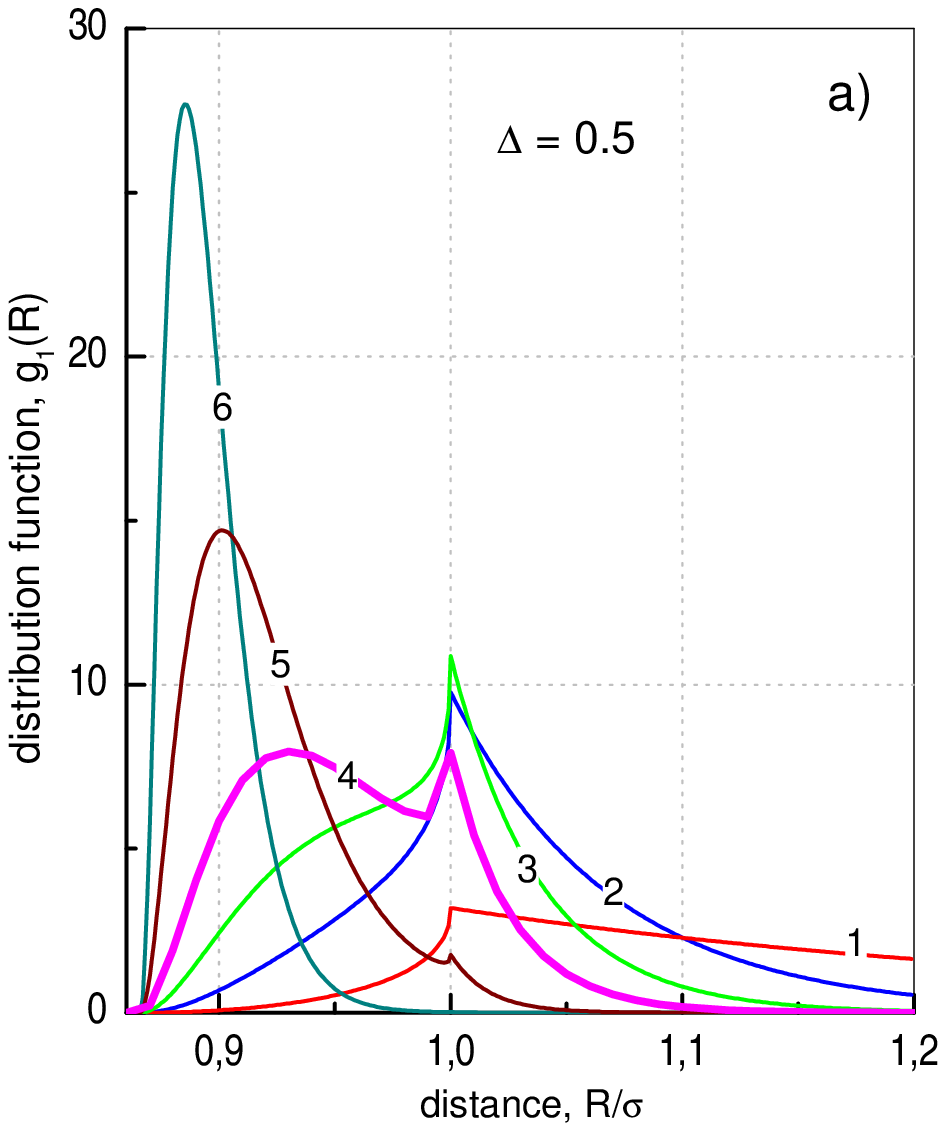} %
\includegraphics[width=0.475\textwidth]{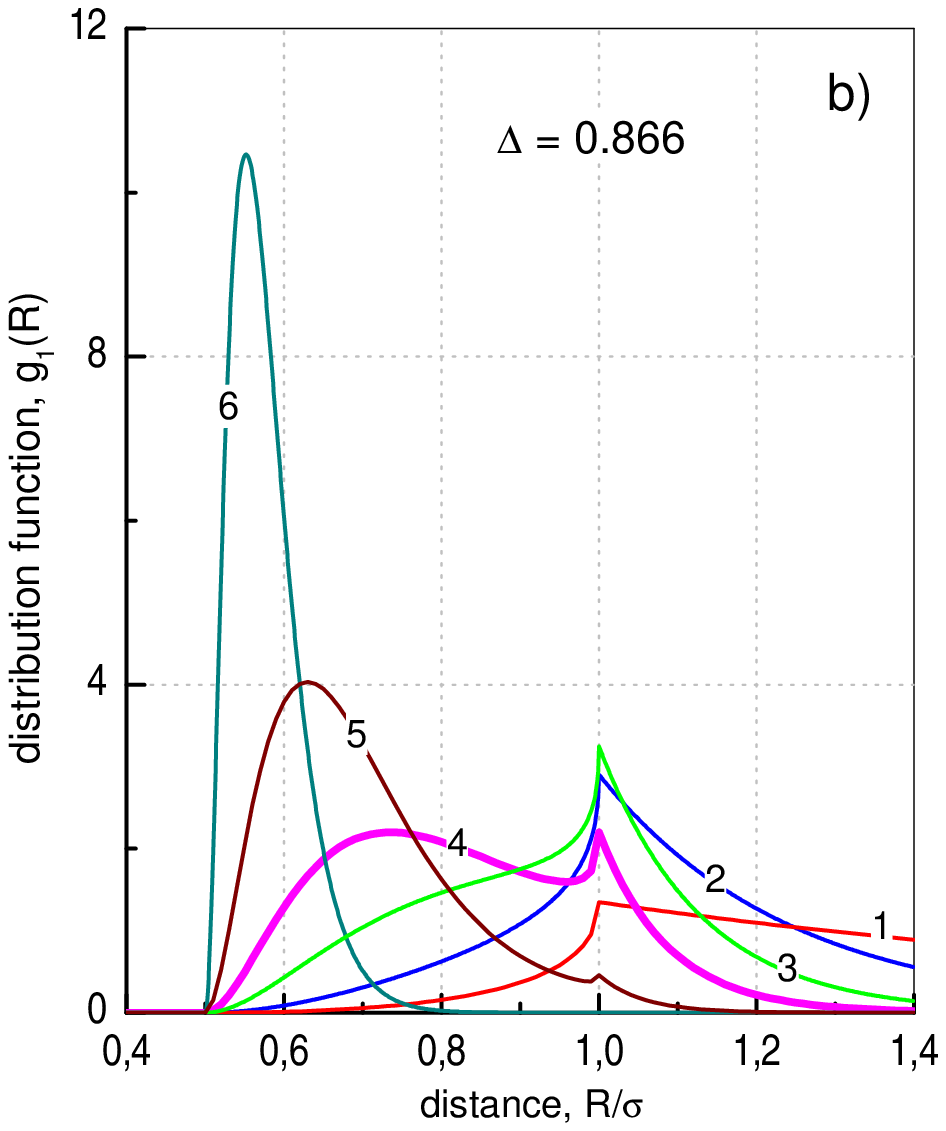}
\caption{Distribution function $\,g_{1}(R)\,$ of the distance $\,R\,$
between next neighbor disks for the pore width: $\Delta =0.5$ and different
linear density $\,\protect\rho _{N}\,$: 1 - $\protect\rho =0.8,$ 2 - 1, 3 -
1.053, 4 -1.08, 5 - 1.111, 6 - 1.13 in part a) and $\Delta =0.886$ and
different linear density $\,\protect\rho _{N}\,$: 1 - $\protect\rho =0.555,$
2 - 1, 3 - 1.25, 4 -1.428, 5 - 1.613, 6 - 1.818 in part b). The density $\,%
\protect\rho =1.08$ in part a) and density $\,\protect\rho =1.428$ in part
b) shown by the thick solid lines and noted as 4, divide all densities into
those with higher $\,R=1$ peak and those with higher $\,R=l_{N}$ peak.}
\label{g1R}
\end{figure}

The longitudinal pair correlations as functions of the disks' number
difference, $g_{2}(|n_{2}-n_{1}|),$ have been investigated in detail by the
transfer matrix method \cite%
{Varga,Godfrey2014,Godfrey2015,Robinson,Hu,Comment}. At the same time, the
PDF $g(R)$ as function of the disk separation $R$ for given density cannot
be directly obtained by this method.  Formula (\ref{PDF}) considerably
simplifies its calculation and enables one to get its systematic
understanding by means of the direct calculation. We numerically obtained
the PDF $g(R)$ (\ref{PDF}) by performing the integration in the formula (\ref%
{Jn}) directly for different pore widths $\Delta $ and linear densities $%
\rho .$ Contrary to our suggestion in \cite{WE} and in line with the results
of \cite{Comment}, our findings on the longitudinal correlations show an
exponential decay for all pore widths and densities. To combine both width
and density effects, we fixed the ratio $\rho /\rho _{\max }$ of the actual
density $\rho $ to the maximum density $\rho _{\max }(\Delta )$ for a given
pore width $\Delta ,$ and then found the correlation lengths for different $%
\Delta $ in the total range of the single-file widths, $0\leq \Delta \leq 
\sqrt{3}/2\approx 0.866\,$, Fig.3\thinspace . For a given $\Delta \,$ the
maximum density is $\,\rho _{\max }(\Delta )=1/\sigma _{m}(\Delta )=1/\sqrt{%
1-\Delta ^{2}}\,$. It follows that as $\Delta $ runs from $0$ to 0.866, the
actual density $\rho =$ $(\rho /\rho _{\max })/\sqrt{1-\Delta ^{2}}$
monotonically increases from $0$ to $1.\,\allowbreak 154\,7(\rho /\rho
_{\max })$. In particular, for the same $\Delta \,$, the actual $\rho \,$ is
higher for higher $\rho /\rho _{\max }\,$. The results for $\rho /\rho
_{\max }=0.866$, 0.9539 and 0.9875 are presented in Fig.3\thinspace .

\begin{figure}[tbp]
\includegraphics[width=0.75\textwidth]{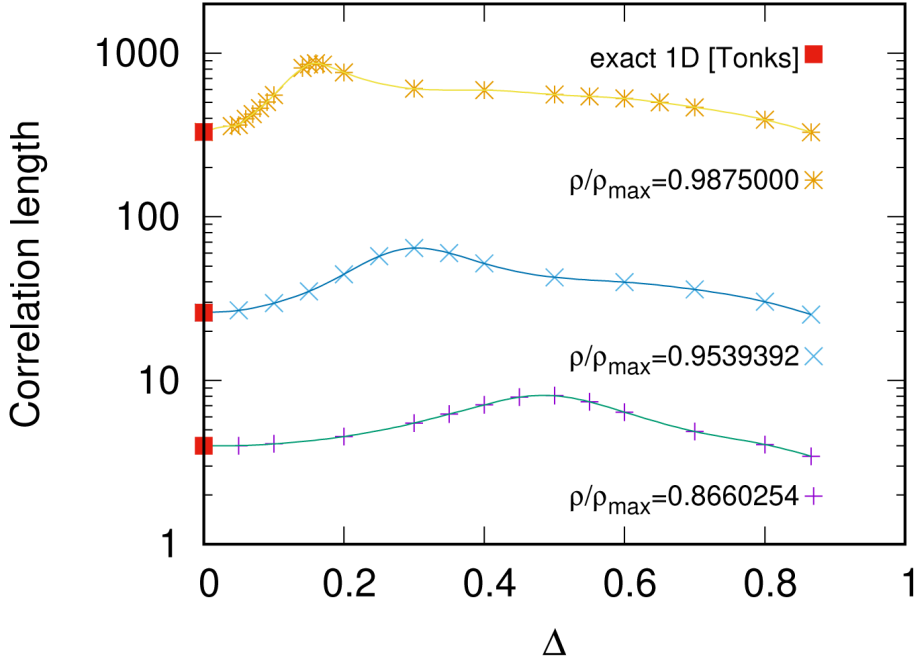}
\caption{Correlation length for the pair correlations $\,g(R)-1\,$ as a
function of pore width $\,\Delta \,$ for the fixed ratio $\protect\rho /%
\protect\rho_{\max }$ (linear density)/(maximum density $1/\protect\sigma %
_{m}(\Delta )$ for given $\Delta ).$ The three curves correspond to the
three different $\protect\rho /\protect\rho _{\max }$ indicated in the
figure. All the three maxima appear exactly at $\protect\rho =1\,$.}
\label{CorL}
\end{figure}

First, it is seen that, for the same $\Delta\,$ the correlation length is
larger for a higher density. Second, as the width approaches zero, the
correlation length tends to the value obtained for the 1D Tonks gas from $%
g_{1D}(x)\,$, Eq.~(\ref{g1D}). Third, the width and density monotonically
grow along the curves in Fig.3\,. It is seen however that the correlation
length does not monotonically increase as both the width and density do:
there is a maximum at each of the three curves. But the most interesting
observation is that all these maxima occur exactly at the density $\rho =1$
when a pore length $R$ equals to the disk diameter $\,d\,$ is occupied by
exactly one disk. This is another peculiarity of these density and disks'
separation indicated above.

\section{Orientational order}

\subsection{Relation with the transfer matrix method}

The orientational order is usually accounted for by the PDF $\left\langle
y_{i}y_{i+n}\right\rangle $ that describes correlation between the
transverse coordinates of disk $i$ and disk $i+n$ \cite%
{Varga,Godfrey2014,Godfrey2015,Robinson,Hu,Comment}. To find this average
from the PF, we use the general formula for the canonical PF in terms of the
integrals over the transverse coordinates of all disks \cite{JCP2020}%
\thinspace : 
\begin{equation}
\left\langle y_{i}y_{i+n}\right\rangle =\int \frac{dl^{\prime }}{Z_{N}}%
\int\limits_{-\infty }^{\infty }d\alpha e^{\displaystyle{N[i\alpha l^{\prime
}+\ln (l-l^{\prime })]}}\prod\limits_{k=1}^{N}\int\limits_{-\Delta
/2}^{\Delta /2}dy_{k}y_{1}y_{n+1}e^{\displaystyle{-i\alpha \sigma
(y_{k}-y_{k-1})}}\,,  \label{yy1}
\end{equation}%
where $\sigma (y_{k}-y_{k-1})$ is defined in Eq.~(\ref{sigma}). Performing
the $\alpha $ and $l^{\prime }$ integration by the steepest descent method,
one obtains\thinspace : 
\begin{equation}
\left\langle y_{i}y_{i+n}\right\rangle =\frac{\int\limits_{-\Delta
/2}^{\Delta /2}dydy^{\prime }yy^{\prime }\left( K^{n}\right) _{yy^{\prime }}%
}{\int\limits_{-\Delta /2}^{\Delta /2}dydy^{\prime }\left( K^{n}\right)
_{yy^{\prime }}}\,,  \label{yy2}
\end{equation}%
where 
\begin{equation}
K_{yy^{\prime }}=e^{\displaystyle{-a_{N}\sigma (y-y^{\prime })}}\,.
\label{K}
\end{equation}%
We see that finding $\left\langle y_{i}y_{i+n}\right\rangle $ from the
canonical PF reduces to the TMM with the TM $K_{yy^{\prime }}$ similar to
that of Kofke and Post \cite{Kofke}. This method has been well established
and the results are well known \cite%
{Varga,Godfrey2014,Godfrey2015,Robinson,Hu,Comment}: For a large number
distance $\,n\,$ between disks, the correlation decays exponentially, i.e., $%
\,g_{yy}(n)=\left\langle y_{i}y_{i+n}\right\rangle -\left\langle
y\right\rangle ^{2}\sim (-1)^{n}\exp (-n/\xi )$ with the correlation length $%
\xi =1/\ln (\lambda _{0}/|\lambda _{1}|)$ where $\lambda _{0}$ and $\lambda
_{1}$ are the leading and subleading eigenvalues of the operator $%
K_{yy^{\prime }}\,$. There is however certain difference between finding the
transverse correlation function from our canonical PF and that from the
isobaric PF upon which the transfer matrix method of Refs.~\cite%
{Kofke,Varga,Godfrey2014,Godfrey2015,Robinson,Hu,Comment} is based. First,
while in our canonical $NLT$ ensemble the quantity $a_{N}$ entering the
operator $K_{yy^{\prime }}$ (\ref{K}) is known before performing the
transfer matrix calculations, its counterpart in the isobaric $NPT$
ensemble, the longitudinal pressure $P_{L},$ has yet to be computed by means
of a nontrivial numerical procedure after the transfer matrix analysis has
been performed. There is also a difference between the $y$ integration
domain in the canonical PF \cite{JCP2020} and the $y$ integration domain in
the isobaric PF obtained after Laplace's transformation in \cite{Kofke},
which was addressed in detail in Appendix to Ref.\cite{JCP2020}. According
to the Authors of Ref.\cite{Comment}, the ensuing difference disappears
because the $NLT$ and $NPT$ ensembles are equivalent in the thermodynamic
limit. We believe that the difference between the predictions of the two
theories can indeed become negligible in an infinite q1D system, but
resorting to the ensemble equivalence in this particular case does not seem
to be justified as this implies that, in particular, the $y$ integration
domains in both PFs are mathematically equivalent. The reason for the
apparent (not strictly proved) equivalence was outlined in \cite{JCP2020}
along with the pointing to the formal inconsistency (missing theta function)
in the Laplace transform results. Namely, the steepest descent integration
results can coincide because the total contribution to the PF integrals
comes from the single maximum (saddle) point which can lie within both not
strictly coinciding integration domains in the $NLT$ and $NPT$ PFs. Thus the
intermediate inconsistency remains but, the result is correct if the saddle
point belongs to the common fraction of the two domains. We saw however that
finding the longitudinal correlation function $\,g(R)\,$ involves computing
PFs of a finite system for which the saddle point does not give the total
contribution. For a finite size system the above difference in the
integration domains can become substantial. For instance, the analysis shows
that for $\Delta =0.5$ and $\rho \sim 1.1\,$, the saddle point method starts
working only for $\,N>10^{6}\,$. Thus, for a q1D HD system with a finite
number of disks, the predictions of PF in the form of Eq.~(\ref{Z1}) or (\ref%
{ZnR}) and of the transfer matrix approach can differ.

It is instructive to compare the longitudinal force $a_{N}$ computed in the
thermodynamic limit from the exact equation (\ref{E}) of the canonical
ensemble \cite{JCP2020} and by the transfer matrix method in \cite{Comment}
for $\Delta =0.5$ and some $\rho\,$ (see Fig.1 of \cite{Comment}). The
results are presented in Table 1\,. 
\begin{table}[h!]
\caption{Longitudinal force $a_{N}$ computed in the thermodynamic limit from
Eq.~(\protect\ref{E}) of the canonical ensemble \protect\cite{JCP2020} and
by the transfer matrix method \protect\cite{Comment} for por width $\Delta
=0.5$ and set of linear densities $\protect\rho\,$. }%
\begin{tabular}{|c|cc|c|}
\hline\hline
$\rho $ & $\beta F_{L}$ &  & $a_{N}$ \\ 
& transfer matrix~\cite{Comment} &  & Eq.~(\ref{E}) [Eq.~(24)~\cite{JCP2020}]
\\ \hline
0.9091 & 7.694 &  & 6.48 \\ 
1.0101 & 20.76 &  & 16.09 \\ 
1.0526 & 29.84 &  & 26.045 \\ 
1.1111 & 61.30 &  & 62.2 \\ 
1.1400 & 181.9 &  & 181.61 \\ \hline\hline
\end{tabular}%
\end{table}

We see some difference between the two predictions. We do not know for sure
what makes this difference, but the most plausible reason is that it comes
from the nontrivial numerical procedures of first planting the system
configurations and then finding the pressure in the $NPT$ ensemble \cite%
{Comment}. If so, however, then one would expect the difference to grow for
higher densities, but, contrary to that, the table shows the opposite trend.
We believe that drawing attention to this difference is worth, but the data
is insufficient for any reasonable conclusion.

\section{Macroscopic orientational order parameter.}

The quantity $\left\langle y_{i}y_{i+n}\right\rangle $ is microscopic as it
measures the correlation between two fixed disks. As in the case of
longitudinal correlations, it is also important to describe certain
correlations, related to the orientational order, as a function of distance
between two point along the pore. Such correlations are expected to be of a
more macroscopic level as they must include contributions from disks with
different $n$ similar to the case of $g(R)$, Eq.~(\ref{PDF}). We know that
there is no long-range translational order in a q1D HD system, hence the orientational
order at best can be of a liquid crystalline type. But this kind of order is
described in terms of a macroscopic order parameter such as, e.g., director
in a nematic liquid crystal. The director is a macroscopic local quantity
that can be obtained by averaging molecular orientations over a small (but
macroscopic) volume centered at the chosen location, and the orientational
order is characterized by the spatial dependence of correlations between the
directors at two separated locations. Thus, the orientational correlation
function in a nematic liquid crystal also has a macroscopic nature. We now
argue that there is a macroscopic order parameter pertinent to the zigzag
structure in a q1D HD system.

First, we would like to present some arguments showing that the microscopic
correlation function $g_{yy^{\prime }}(n)$ can result in an ambiguous
interpretation of the orientational order. The decay of $g_{yy^{\prime }}(n)$
has been attributed to the appearance of windowlike defects in the zigzag
arrangement \cite{Saika,Godfrey2014,Robinson,JCP2020,WE} and the correlation
length $\xi $ has been related to the mean separation of such defects \cite%
{Saika,Godfrey2014,Robinson}. Such a defect implies a pair disks separated
by the horizontal contact distance equal to the disk diameter ($\sigma =1)$
which, in dense zigzags, with a high probability implies that the disks'
bond is along the pore and thus the up-down-up-down... order is violated and
changed to, e.g., up-down-down-up... For a density very near close packing, $%
g_{yy}(n)$ must show a long sequence $-1,1,-1,1,-1,1,-1...$ until a defect
at some $n_{d}$ changes it into $-1,1,-1,-1,1,-1,1...$ What happens with $%
g_{yy}$ can be described as follows: Before the defect, $g_{yy}($odd $n)=-1$
and $g_{yy}($even $n)=1,$ but after the defect, when $n>n_{d}$, $g_{yy}($odd 
$n)=1$ and $g_{yy}($even $n)=-1.$ We see that all that happened with the
zigzag is the relative $\pi $ phase shift between long highly ordered zigzag
arrays$,$ which does not seem to smear overall zigzag order. Moreover, after
a second defect at some higher $n_{d}^{\prime },$ the phase restores by
another $\pi $ shift so that for $n>n_{d}^{\prime }$ the zigzag is not
shifted and the microscopic correlation function $g_{yy^{\prime }}(n)$ is
the same as in the beginning.

Next, by analogy with $g(R)$ (\ref{PDF}), consider the spatial orientational
correlation function obtained by summing contributions from disks with
different $n$ that at some instant appear at a separation $R$ from the disk
with $n=0$ at $R=0.$ For a high density, such contributions would be sign
alternating and thus giving a very small value of the correlation at $R$
while the zigzag order is actually very high. This consideration suggests
that a macroscopic order parameter has to ignore defect induced phase shift,
but instead incorporate their presence as its magnitude decrease. Below we
propose such a local macroscopic orientational order parameter.

Here we first introduce the orientational order parameter $\,\psi (\rho)\,$
that does not account for the disk separation order along the pore. Let $\,y_{1}\,$,  $\,y_{2}\,$ and $\,y_{3}\,$ be the vertical coordinates of three neighboring
disks 1, 2 and 3 that range from $\,-\Delta /2\,$ to $\,\Delta /2\,$. We will need
the DF $\,f_{y}(y)\,$ of the $\,y\,$ coordinate of a single disk and DF $\,f_{\delta
y}(\delta y_{21})\,$ of the difference $\,\delta y_{21}=y_{2}-y_{1}\,$ for two
neighbors with the transverse coordinates $\,y_{2}\,$ and $\,y_{1}\,$ derived in 
\cite{JCP2020}\,; these two DFs will be given below. We define the
local orientational order parameter as the mean value of the product $\,\left\langle
f_{y}(y_{1})\delta y_{21}\delta y_{32}\right\rangle\,$ : 
\begin{eqnarray}
\psi (\rho ) &=&\frac{-1}{\Delta ^{2}Z_{\psi }}\int\limits_{-\Delta
/2}^{\Delta /2}dy_{1}f_{y}(y_{1})\int\limits_{-\Delta /2}^{\Delta
/2}dy_{2}f_{\delta y}(y_{2}-y_{1})(y_{2}-y_{1})\int\limits_{-\Delta
/2}^{\Delta /2}dy_{3}f_{\delta y}(y_{3}-y_{2})(y_{3}-y_{2})\,,  \notag \\
&&  \label{psi}
\end{eqnarray}
where 
\begin{equation}
Z_{\psi }=\int\limits_{-\Delta /2}^{\Delta
/2}dy_{1}f_{y}(y_{1})\int\limits_{-\Delta /2}^{\Delta
/2}dy_{2}\int\limits_{-\Delta /2}^{\Delta /2}dy_{3}f_{\delta
y}(y_{2}-y_{1})f_{\delta y}(y_{3}-y_{2})\,.
\end{equation}
For a densely packed zigzag, $\,-\left\langle \delta y_{21}\delta
y_{32}\right\rangle =-\left[ \Delta \times (-\Delta )\right] =\Delta ^{2}\,$
so that $\psi (\rho _{dp})=1$ whereas for low density this is expected to
tend to zero. The DFs $\,f_{y}(y)\,$ and $\,f_{\delta y}(\delta y)\,$ can be
expressed via the TM operator $\,K_{yy^{\prime }}$ \cite{f y}\,: 
\begin{eqnarray}
f_{y}(y) &=&\left( \frac{\displaystyle{\int_{-\Delta /2}^{\Delta
/2}K_{yy^{\prime }}dy^{\prime }}}{\displaystyle{\int_{-\Delta /2}^{\Delta
/2}K_{yy^{\prime }}dydy^{\prime }}}\right) ^{2},  \label{f_y} \\
&&  \notag \\
f_{\delta y}(y-y^{\prime }) &=&\frac{K_{yy^{\prime }}}{\displaystyle{%
\int_{-\Delta }^{\Delta }K_{yy^{\prime }}d(y-y^{\prime })}}\,.  \label{f_del}
\end{eqnarray}
The pressure $a_{N}$ is the solution of the Eq.~(\ref{E}) so that for given
pore width $\Delta \,$, parameter $\psi \,$ is a function on $\,l_{N}=1/\rho\,$. The order parameter $\,\psi\,$ describes the quality of the single
structural zigzag element $\, y_{1}-y_{2}-y_{3}\,$ consisting of two bonds; it
ignores the defect induced phase shift as it is quadratic in $\,\delta y\,$ and
positive for a high order;  if there is a defect then $\,\delta y\sim 0\,$ and
the magnitude of $\,\psi\,$ decreases; because of its quadratic form it does
not vanish upon integration over a macroscopic volume and thus can be used
as a macroscopic quantity. Figure~4 shows that it indeed decreases from 1 to
0 as the density decreases and the zigzag order weakens.

\begin{figure}[tbp]
\includegraphics[width=0.45\textwidth]{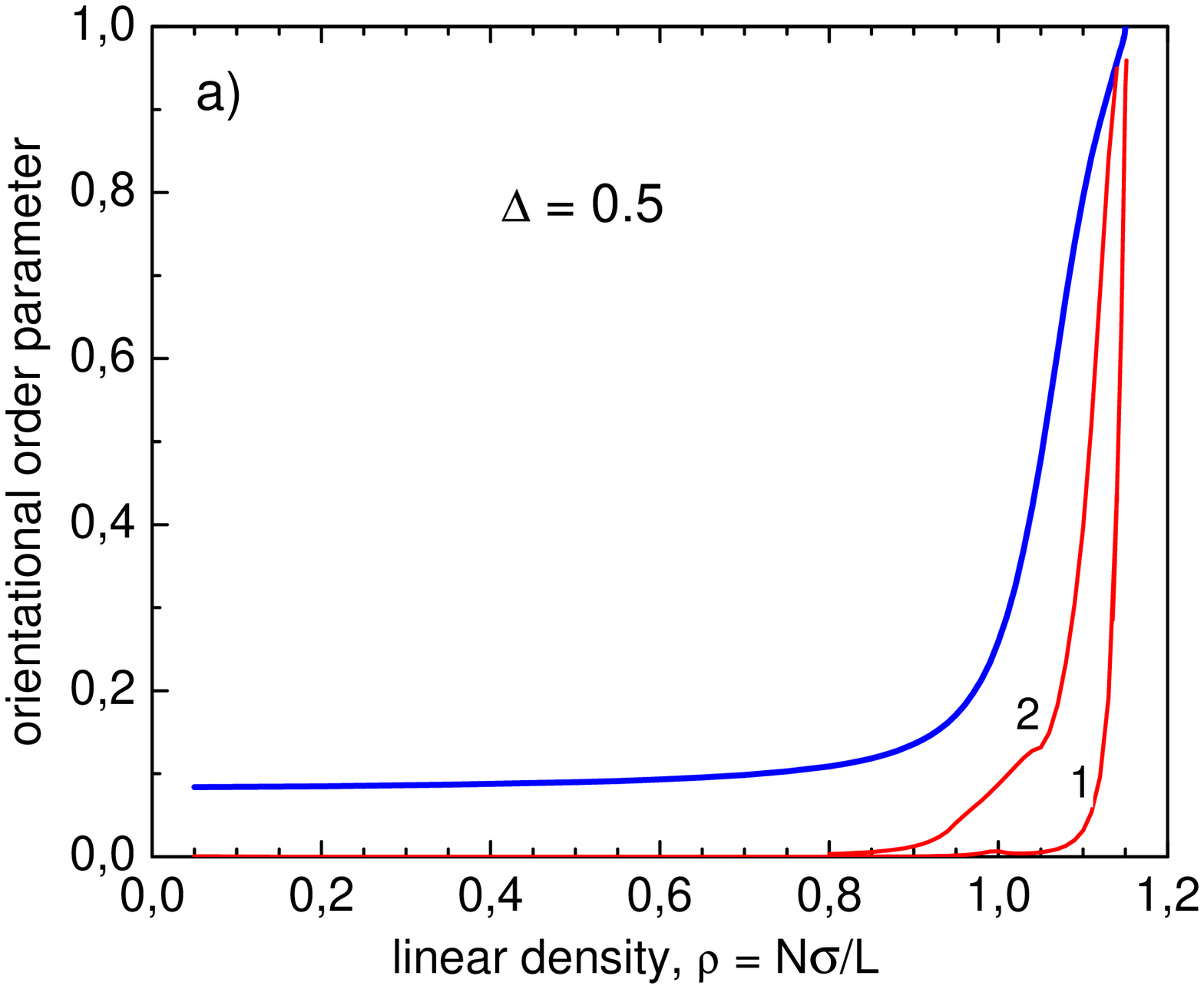} %
\includegraphics[width=0.45\textwidth]{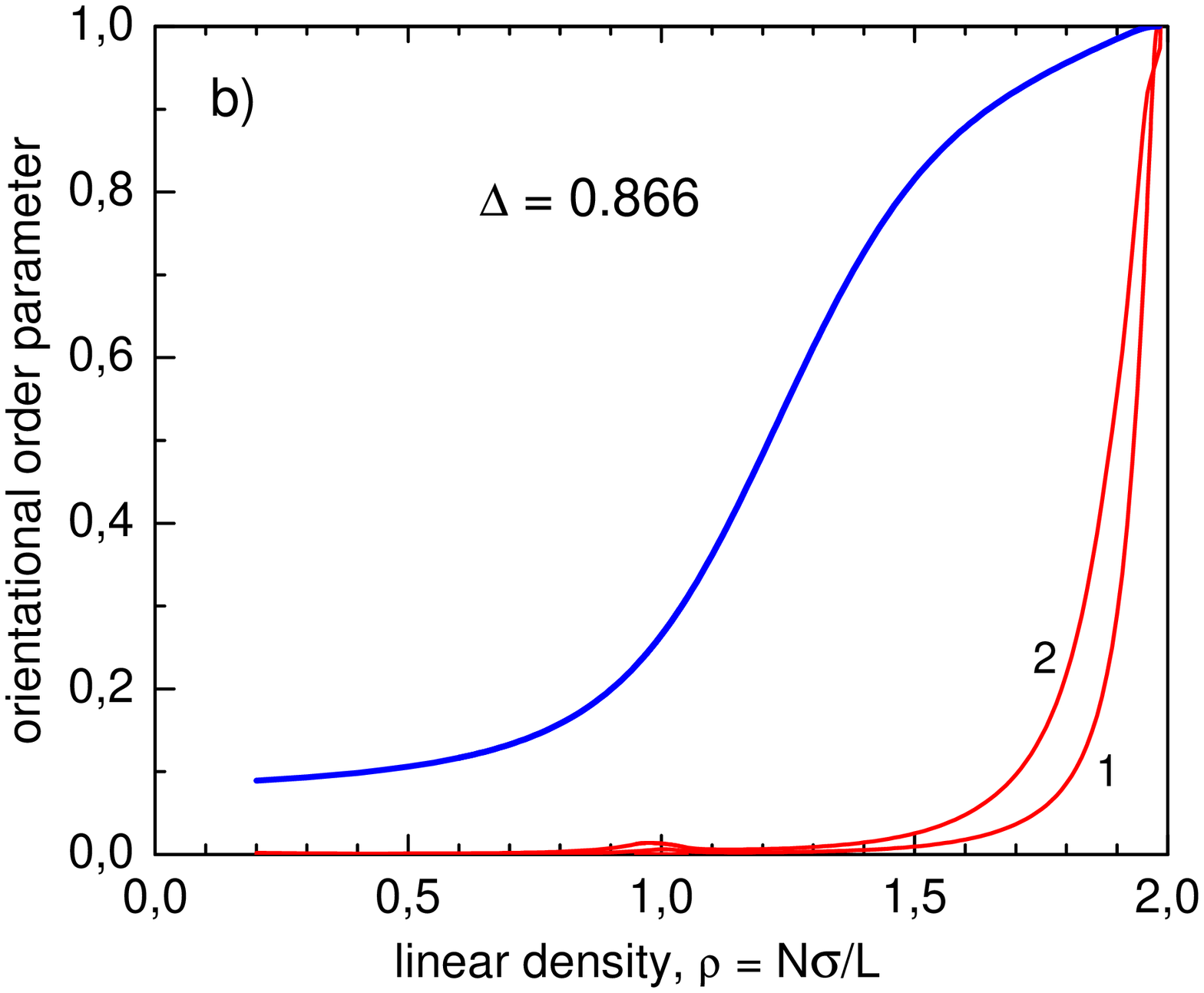}
\caption{Orientational order parameter $\,\protect\psi (\protect\rho)\,$ (1)
and the total $\,$order parameter $\Psi (\protect\rho )$ for $\protect\delta %
l=0.1l_{N}$ (2)$,0.05l_{N}$ (3), and $0.01l_{N}$ (4). Part a) $\Delta =0.5\,$%
, part b) $\Delta =0.866\,$.}
\label{PsiRho}
\end{figure}

In the parameter $\psi(\rho)\,$, the perfect order is associated with both $\,\delta y_{21}\,$ and $\,\delta y_{32}\,$ being of maximum magnitude $\,\Delta \,$.
This however is not sufficient because in the perfect zigzag in addition
disks' spacing along the pore must be equal to its average value $%
l_{N}=1/\rho \,$. To account for deviations from this perfect spacing, one
can use the probability that the spacing is concentrated around $R=l_{N}\,$.
The quantity $\,g_{1}(R)\,$, Eq.~(\ref{gg1}), gives the probability density
rather than probability, and can be thus unbounded [for close packing $%
g_{1}(R=l_{N})=\delta (R-l_{N})]\,$. To incorporate $g_{1}(R=l_{N})$ in the
order parameter we can convert it into probability of some configuration in
the vicinity of the ideal geometry $R=l_{N}$ that we choose to represent a
good longitudinal zigzag pattern. For example, we can describe the
horizontal aspect of zigzag's quality by the probability $w(l_{N},\delta l)$
that the horizontal distance between disks falls into the interval $%
[l_{N}-\delta l,l_{N}+\delta l]$ where$,$ for instance, $\delta l$ cannot be
larger than one percent of $l_{N},$ i.e., $\delta l=\min
(0.01l_{N},l_{N}-\sigma _{m}).$ This construction takes into account that
for high densities the deviation by one per cent from $l_{N}$ cannot be
achieved as the minimum distance between disks' centers, $\sigma _{m},$ in
this case can be larger than $l_{N}(1-0.01)\,$. This $\,w(l_{N},\delta l)\,$
is 
\begin{equation}
w(l_{N},\delta l)=\int_{l_{N}-\delta l}^{l_{N}+\delta l}g_{1}(R)dR\,,
\label{w}
\end{equation}%
with $\,\delta l=\min (0.01l_{N},l_{N}-\sigma _{m})\,$.

For close packing, $w$ is maximum equal to $1$. The improved order parameter
is now 
\begin{equation}
\Psi (\rho )=w^{2}(l_{N},\delta l)\psi (\rho )\,,  \label{PSI}
\end{equation}
where $w^{2}$ reflects that the chosen structural zigzag element consists of
two bonds. This $\Psi (\rho )$ is shown in Fig.4; it is seen that its
behavior between low and high densities is sharper than that of $\,\psi\,$
which is the effect of the spacing order along the pore. \ This parameter is
local in the sense that its average over a finite volume $\Delta L$ is
finite, but it does not depend on the location of this volume in the system.
In other words, 
\begin{equation}
\overline{\Psi }(\rho, \Delta )=\frac{1}{\Delta L}\sum_{\text{all }yyy\text{
in }\Delta L}\Psi = const(\rho, \Delta )\,,  \label{PSI bar}
\end{equation}
where the dependence on the width is shown explicitly. In particular, $%
\,\Delta L\,$ can be the total $\,L\,$, so that macroscopic orientational
order is of long range and is of a liquid crystal type. This is the
consequence of the specific q1D geometry. In 2D crystals, under thermal
fluctuations, the structural lattice pattern (hexagon in the case of
triangular lattice) can turn with respect to the one separated by a distance 
$R,$ and the lager so the larger $R$ is. As a result, the rotation angle
between the two patterns makes them uncorrelated. In the q1D geometry, the
walls prevent zigzag from rotation, the only dephasing consists of the $%
\,\pi\,$ shifts whose density is homogeneous along the system so that their
impacts on the magnitude of $\,\Psi(x)\,$ at different locations $x$ are
similar$,$ the order parameter at different $x$ are correlated and do not
cancel one another in the sum over $\,x\,$.

\section{Discussion}

We derived the formulae for the important PDFs $g(R)$ and $g_{1}(R)$ and
demonstrated that they can be readily used. Apart of that we also made a few
suggestions. First, based on our finding on the correlation lengths, we
suggested that the density $\rho =1$ plays a distinguished role in the
zigzag transformation with density irrespective of the pore width. Second,
we pointed out that windowlike defects can not only destroy, but also
restore the order so that they are not the direct reason for the decay of $%
g_{yy}(n).$ Third, we suggested that in a q1D HD system there is a
macroscopic orientational order parameter that has similarity to the
macroscopic order parameter of a nematic liquid crystal. Here we discuss
these suggestions in a more detail.

Figures 2-4 show that the PDFs $g_{1}(R)$, $g(R),$ and the order parameter $%
\,\Psi(\rho)\,$ have peculiarities at the density $\rho =1$ in the form of
certain peaks or maxima. It has been suggested that the peak at the
distribution of next neighbors, Fig.2, is related to the tendency of the
system to produce windowlike defects to increase the entropy as such a
defect enables disk's travel across the pore \cite%
{Saika,Godfrey2014,Robinson,JCP2020,WE}. However, our finding that the
correlation length has a maximum exactly at $\rho =1$ for any pore width,
Fig.3, is unexpected and cannot be explained by this idea. At higher
densities, the peak at $\rho =1$ is diminishing and the peak at another
distinguished, namely average density $\rho =N/L$ is raising and eventually
dominates the one at $\rho =1$ (the peak at $\rho =1\,$, however small, is
present for any density \cite{JCP2020}). As the peak of $g_{1}(R)$ at the
reciprocal density is definitely related to the longitudinal component of
the zigzag order, it is natural to connect the peak at $\rho =1$, at least
partially, to the nascent longitudinal ordering, too. In the light of this
idea, the maxima of the correlation length become reminiscent of the
correlation length increase at a phase transition. Of course, there is no
transition at $\rho =1,$ but the echo thereof seems to show up. We may then
speculate that at $\rho =1,$ this echo of a nascent zigzag order with a low
orientational transverse magnitude and higher longitudinal component is
somehow related to the increase of the correlation length at $\rho =1.$

Now consider how the windowlike defects and the order decay are related. To
this end we consider the formula (\ref{yy2}) as a stochastic process
(Markov's chain) with the probability matrix $f_{\delta y}(y_{i+1}-y_{i})$,
Eq.~(\ref{f_del}). This $\,f_{\delta y}\,$ is the probability density that a disk $i$
with the transverse coordinate $y_{i}$ has a next neighbor disk $i+1$ with
the coordinate $y_{i+1}.$ In the perfect zigzag, $f_{\delta
y}(y_{i+1}-y_{i})=\delta (y_{i+1}-y_{i}-\Delta )$ showing that all disks
stay at the walls: a disk at one wall can have a next neighbor only at the
opposite wall. Iterating this process we will find that all disks stay
either at one wall or at another. If$,$ however, $\rho $ is below the dense
packing value, disks can stay at any distance from the walls. Of course, for
a high density, the probability that a disk with $y_{i}$ close to one wall
can have a next neighbor with $y_{i+1}$ close to another wall is
considerably higher than the probability that, e.g., the next neighbor stay
at the same wall, i.e., $y_{i}=y_{i+1},$ but it is nonzero. Continuing the
iterations, we will find progressively more and more uniform probability of
the coordinate $y_{n+1}$ of the $n$-th neighbor of disk $1$. As a result,
for any initial $y_{i},$ the $y_{i+n}$ integral is getting less an less
until for infinite $n$ the distribution ($K^{n})_{y_{i}y_{i+n}}$ becomes the
constant $1/\Delta $ and $\int_{-\Delta /2}^{\Delta
/2}dy_{i+n}(K^{n})_{y_{i}y_{i+n}}$ vanishes. In other words, the reason for
the correlation decay is that the transition probability $K_{yy^{\prime }}$
is finite for any $y$ and $y^{\prime }$ so that the mapping $K^{n}$ is
ergodic. It is clear that the closer the function $f_{\delta
y}(y_{i+1}-y_{i})$ to the delta function, the slower the iterations converge
to the uniform distribution and the larger the correlation length. Now we
note that the above situation with $y_{i}=y_{i+1}$ with a nonzero
probability is the probability of a windowlike defect. We see that defects
in themselves are the consequences of the finite values of $K_{yy^{\prime
}}. $ Thus, the defects are connected with the correlation decay, but they
are the effect of the decay rather that its cause. Of course, whatever is
the causality, the connection between the decay rate and defect density does
exists and can be used for finding one quantity if the other is known.

Now let us justify the introduction of the macroscopic orientational order
parameter making a parallel with the standard idea of the nematic order. As
in our q1D HD system, a translation order in a nematic liquid crystal does
not exist. If the orientational order is not perfect (scalar order parameter
is below unity), then the PDF for neighboring molecules is maximum for their
parallel alignment, but is nonzero for any their mutual orientation. If now
one draws a line, innumerate molecules staying along this line and construct
the chain with the PDF as a transition probability, the result will be
similar to that in our q1D system, i.e., the correlations will be decaying
along the line. Moreover, there are many impurities and ions in a nematic
liquid crystal that do not facilitate the orientational order and can play
the role of defects if appear on the chosen line. The macroscopic order
nevertheless does exist as the molecular orientations averaged over a
macroscopic volume give the macroscopic director while the impurities and
the imperfect PDF make the scalar order parameter (the degree of order)
lower. The macroscopic PDF considered in statistical physics of a nematic
liquid crystal is that between the macroscopic directors and is governed by
the macroscopic elastic free energy \cite{Landau5}. Of course, the order
parameter we introduced here does not play the role so fundamental as the
director in a nematic liquid crystal, but we believe that it is reasonable
to consider macroscopic order if it can be physically justified in a q1D
system of hard disks. Note that the introduced order parameter is apolar as
it ignores the $\pi $ shifts and in this sense is similar to the apolar
nematic director.

In conclusion, we believe that the results of this paper complement recent
studies of low dimension models and advance their analytical tools and our
understanding of these systems.

\section{Data Availability Statement}

Data sharing is not applicable to this article as no new data were created
or analyzed in this study.

\textit{Acknowledgment.}%
\begin{acknowledgements}
V.M.P. is grateful for financial support and hospitality of the Polish Academy of Sciences and the Center for Theoretical Physics PAS.
\end{acknowledgements}

\appendix{}{}

\end{document}